\documentclass[10pt,conference]{IEEEtran}

\usepackage{booktabs}
\usepackage{amsmath}
\usepackage{amsthm}
\usepackage{multirow}
\usepackage{color,cite}
\usepackage{subfigure}
\usepackage{mathrsfs}
\usepackage{amssymb}
\usepackage{bm}
\usepackage{graphicx}
\usepackage{amsfonts}
\usepackage{algorithm}
\usepackage{algorithmic}
\usepackage{epstopdf}

\newcommand{\tabincell}[2]{\begin{tabular}{@{}#1@{}}#2\end{tabular}}

\begin{document}

\bibliographystyle{IEEEtran}

\title{\vspace{-2mm}\huge{Improving the Performance of R17 Type-II Codebook \\ with Deep Learning}\vspace{-4mm}}

\author{\normalsize
\vspace{-4.2mm}
\IEEEauthorblockN{Ke Ma$^{\rm 1}$, Yiliang Sang$^{\rm 1}$, Yang Ming$^{\rm 1, \rm 2}$, Jin Lian$^{\rm 3}$, Chang Tian$^{\rm 3}$, Zhaocheng Wang$^{\rm 1, \rm2}$, \emph{Fellow, IEEE}} \\
\IEEEauthorblockA{$^{\rm 1}$Department of Electronic Engineering, Tsinghua University, Beijing 100084, China\\
 $^{\rm 2}$Shenzhen International Graduate School, Tsinghua University, Shenzhen 518055, China \\
 $^{\rm 3}$Huawei Technologies Company, Ltd., Beijing 100095, China \\
 \{ma-k19, sangyl19, mingy20\}@mails.tsinghua.edu.cn, \{lianjin7,tianchang7\}@huawei.com, zcwang@tsinghua.edu.cn}
\vspace{-11.3mm}}


\maketitle

\begin{abstract}
The Type-II codebook in Release 17 (R17) exploits the angular-delay-domain partial reciprocity between uplink and downlink channels to select part of angular-delay-domain ports for measuring and feeding back the downlink channel state information (CSI), where the performance of existing deep learning enhanced CSI feedback methods is limited due to the deficiency of sparse structures.
To address this issue, we propose two new perspectives of adopting deep learning to improve the R17 Type-II codebook.
Firstly, considering the low signal-to-noise ratio of uplink channels, deep learning is utilized to accurately select the dominant angular-delay-domain ports, where the focal loss is harnessed to solve the class imbalance problem.
Secondly, we propose to adopt deep learning to reconstruct the downlink CSI based on the feedback of the R17 Type-II codebook at the base station, where the information of sparse structures can be effectively leveraged.
Besides, a weighted shortcut module is designed to facilitate the accurate reconstruction.
Simulation results demonstrate that our proposed methods could improve the sum rate performance compared with its traditional R17 Type-II codebook and deep learning benchmarks.
\end{abstract}

\IEEEpeerreviewmaketitle

\vspace{-1mm}
\section{Introduction}\label{S1}

Benefitting from the high spectral efficiency, massive multiple-input multiple-output (MIMO) has been regarded as one of the fundamental technologies in the fifth-generation (5G) wireless communications and beyond.
To support efficient precoding, the base station (BS) needs to acquire accurate downlink channel state information (CSI) of each user equipment (UE).
For frequency division duplex (FDD) systems, the CSI acquisition mainly relies on the CSI feedback from UEs, whereas the massive number of antennas could lead to huge feedback overhead \cite{ref1}.


Fortunately, since CSI in massive MIMO systems generally enjoys the sparse property in both angular and delay domains, compressed sensing (CS) based methods could obtain the sufficiently accurate CSI estimates based on a small number of sparse bases \cite{ref2}.
Moreover, the existing studies demonstrated the angular-delay-domain partial reciprocity in FDD systems, i.e., both the uplink and downlink CSI have similar power properties in angular and delay domains \cite{ref5}, which has inspired many works to select the strongest angular-delay-domain ports according to the uplink CSI for downlink channel measurements \cite{ref5, ref6}, so that the downlink pilot overhead can be reduced.
However, the limited transmit power of UE usually leads to low signal-to-noise ratios (SNRs) of uplink CSI, making it hard to accurately select the dominant ports.

In spite of widespread applications, the CS based methods suffer from a crucial drawback that they heavily rely on pre-determined sparse assumptions, whose performance may degrade severely if the practical channel is not consistent with these assumptions.
To address this issue, deep learning has been introduced into CSI feedback to adaptively capture the channel sparsity for enhancing the feedback performance.
Specifically, the auto-encoder (AE) has been broadly applied for feeding back the full CSI matrix \cite{ref7, ref8, ref9, ref10, ref11}.
Besides, by utilizing the angular-delay-domain partial reciprocity, the uplink CSI information was fused in the AE to relieve the burden of UE feeback \cite{ref16}.


Although the existing studies mainly concentrate on the feedback of the full CSI matrix, Third Generation Partnership Project (3GPP) standardized the CSI eigenvector feedback \cite{ref18}, where the CSI eigenvectors of multiple subbands in orthogonal frequency division multiplexing (OFDM) systems are jointly compressed and fed back.
Consequently, exploring efficient CSI eigenvector feedback with deep learning has drawn broad attention.
In \cite{ref19}, EVCsiNet was constructed to exploit the AE architecture for feeding back the CSI eigenvectors.
Besides, the work \cite{ref20} proposed MixerNet to fuse spatial and frequency features for improving the eigenvector feedback accuracy, while the dual-polarized antenna structure was exploited for performance enhancement in \cite{ref21}.
Moreover, the bi-directional long short-term memory network was adopted to accurately model the eigenvector variations among different subbands \cite{ref22}.

However, the above methods for CSI eigenvector feedback focus on improving the performance of the Release 16 (R16) Type-II codebook, which needs to measure the whole downlink CSI eigenvectors for choosing the sparse bases and imposes considerable downlink pilot overhead.
To solve this problem, the Release 17 (R17) Type-II codebook leverages the angular-delay-domain partial reciprocity to select the dominant angular-delay-domain ports according to the uplink CSI and only measures the corresponding channel coefficients, so that the downlink pilot overhead can be significantly reduced \cite{ref18}.
In this scenario, different UEs usually select various ports for downlink channel measurements, making the sparse structure of the feedback coefficient vector vary for diverse UEs.
Unfortunately, UE is not aware of its corresponding ports, hence the stable sparse structure of the feedback vector is severely destroyed, which seriously restricts the performance of deep learning based CSI compression at UE side \cite{ref24, ref25}.

In contrast to directly adopting deep learning for CSI feedback, we propose two new perspectives to improve the feedback performance of the R17 Type-II codebook.
Firstly, considering the low signal-to-noise ratio (SNR) of uplink channel measurement, deep learning is proposed to be utilized to capture the sparse structure in the uplink CSI for accurately selecting the dominant angular-delay-domain ports.
Nevertheless, the selected ports only occupy a small fraction of all ports, and the ports with lower power are difficult to be accurately chosen. To solve this issue, the focal loss is harnessed to adaptively weight the loss function to balance both the positive/negative classes and the simple/hard samples for improving the selection accuracy.
Secondly, we propose to use deep learning to reconstruct the downlink CSI eigenvectors on all angular-delay-domain ports based on the feedback of the R17 Type-II codebook at BS side, where the sparse structure in the angular-delay domain can be sufficiently leveraged.
Since the CSI reconstructed by the Type-II codebook itself is similar to the perfect CSI, we adopt a weighted shortcut module to implement the fine-tuning on the reconstruction input.
Simulation results show that our proposed methods can significantly improve the sum rate compared to its R17 Type-II codebook and traditional deep learning benchmarks.
\vspace{-.7mm}
\section{System Model and R17 Type-II codebook}\label{S2}
\vspace{-.4mm}
\subsection{System Model}\label{S2.1}
\vspace{-.3mm}

Consider a FDD system where one BS equipped with $N_\text{Tx}$ antennas serves $K$ single-antenna UEs.
Assume that the BS is deployed with dual-polarization antennas in the uniform planar array (UPA), and let $N_\text{h}$ and $N_\text{v}$ represent the numbers of dual-polarization antennas at horizonal and vertical directions, so that $N_\text{Tx} = 2 N_\text{h} N_\text{v}$.
Following the 5G standard \cite{ref18}, we further assume that an OFDM system containing $M$ subbands is adopted, where each subband comprises $N_\text{s}$ resource blocks (RBs) and is regarded the granularity of channel estimation, while each RB consists of 12 subcarriers.
Denoting the CSI vector of the $k$-th UE on the $m$-th subband as $\bm{h}_{k,m} \in \mathbb{C}^{N_\text{Tx} \times 1}$, the corresponding CSI matrix on all subbands $\bm{H}_k \in \mathbb{C}^{N_\text{Tx} \times M}$ can be obtained by concatenating the CSI vectors, i.e., $\bm{H}_k = \left[ \bm{h}_{k,1} ~ \bm{h}_{k,2} ~ ... ~ \bm{h}_{k,M} \right]$.

To generate CSI matrices, the clustered delay line channel model defined in 3GPP Specification 38.901 \cite{ref26} is adopted.
Since the uplink and downlink channels in FDD systems enjoy the similar power property in angular and delay domains \cite{ref5}, the 38.901 channel model assumes that the uplink and downlink channels have analogous pathlosses, angles and delays, with slight perturbations depending on the gap of center frequencies.
For convenience, the subscripts UL and DL are adopted to distinguish uplink and downlink channels.

\vspace{-1mm}
\subsection{R17 Type-II Codebook}\label{S2.1}
\vspace{-.3mm}

The R17 Type-II codebook utilizes the angular-delay-domain sparsity and partial reciprocity to reduce the overhead of both downlink measurement and uplink feedback in FDD systems, which comprises four steps as below.
For simplicity, we assume single-antenna UEs in the system model, so that the CSI vector on each subband $\bm{h}_{k,m}$ itself is the CSI eigenvector.
The detailed steps can be found in \cite{ref27}.

\emph{1) Uplink port selection}:
To leverage the angular-delay-domain partial reciprocity, the uplink CSI matrix of the $k$-th UE $\bm{H}_{k,\text{UL}} \in \mathbb{C}^{N_\text{Tx} \times M}$ is firstly transformed into the angular-delay domain as below
\vspace{-1.5mm}
\begin{equation}\label{eq5}
\widetilde{\bm{H}}_{k,\text{UL}} = \bm{W}_\text{A}^\text{H} \bm{H}_{k,\text{UL}} \bm{W}_\text{D},
\vspace{-1.5mm}
\end{equation}
where $\widetilde{\bm{H}}_{k,\text{UL}} \in \mathbb{C}^{N_\text{Tx} \times M}$ is the corresponding angular-delay-domain CSI matrix.
In the left, the angular-domain transform matrix $\bm{W}_\text{A} \in \mathbb{C}^{ N_\text{Tx} \times N_\text{Tx} }$ consists of two identical matrices $\bm{D} \in \mathbb{C}^{ N_\text{h} N_\text{v} \times N_\text{h} N_\text{v} }$ on the diagonal corresponding to dual polarization directions, i.e., $\bm{W}_\text{A} = \text{diag}\left( \bm{D}, \bm{D} \right) $, where the oversampled orthogonal discrete Fourier transform (DFT) bases are adopted in $\bm{D}$.
Similarly, the delay-domain transform matrix $\bm{W}_\text{D} \in \mathbb{C}^{ M \times M }$ in the right comprises $M$ orthogonal DFT bases.
Considering the partial reciprocity, the BS can select the $P$ strongest angular-delay-domain ports for downlink channel measurement, where each port corresponds to a specific angular-delay-domain basis.

\emph{2) Downlink port coefficient measurement}: Based on the angular-delay-domain sparsity, the BS could only transmit the downlink pilots on the selected ports to acquire a sufficiently accurate CSI estimate. Specifically, assume that the $p$-th port for the $k$-th UE adopts the $p^\text{(A)}_{k}$-th column in $\bm{W}_\text{A}$, i.e., $\bm{w}_{\text{A},p^\text{(A)}_{k}} = {\bm{W}_\text{A}}{[:, p^\text{(A)}_{k}]}$, as the angular-domain basis, and adopts the $p^\text{(D)}_{k}$-th column in $\bm{W}_\text{D}$, i.e., $\bm{w}_{\text{D},p^\text{(D)}_{k}} = {\bm{W}_\text{D}}{[:,p^\text{(D)}_{k}]}$, as the delay-domain basis.
Consequently, the corresponding downlink precoding matrix for channel measurement $\bm{\Phi}_{k,p} \in \mathbb{C}^{ N_\text{Tx} \times M }$ can be calculated as $\bm{\Phi}_{k,p} = \bm{w}_{\text{A},p^\text{(A)}_{k}} \bm{w}_{\text{D},p^\text{(D)}_{k}}^\text{H}$.

\emph{3) Uplink port coefficient feedback}: After channel measurement, the two-stage quantization method is applied for CSI compression at UE side. In the first stage, the $Q_\text{w}$-bit wideband amplitude quantization is introduced to indicate the amplitude ratio of two polarization directions.
Then, the amplitude and phase of each narrowband port coefficient are separately compressed by the $Q_\text{n,a}$-bit logarithmic quantization and $Q_\text{n,p}$-bit uniform quantization in the second stage.
After quantization, the quantized port coefficients $\bar{c}_{k,p}, p = 1,2,..., P$ are fed back from the $k$-th UE to BS.

\emph{4) Downlink CSI reconstruction}: Finally, the reconstructed downlink CSI matrix of the $k$-th UE $\bm{H}_{k,\text{DL(TypeII)}}$ can be expressed as below
\vspace{-1.5mm}
\begin{equation}\label{eq8}
\bm{H}_{k,\text{DL(TypeII)}} = \sum_{p=1}^{P} \bar{c}_{k,p} \bm{w}_{\text{A},p^\text{(A)}_{k}} \bm{w}_{\text{D},p^\text{(D)}_{k}}^\text{H}.
\vspace{-1.5mm}
\end{equation}
Considering that the port selection and CSI reconstruction processes are the same for all UEs, the subscript of UE index $k$ is omitted in the following.
\vspace{-1.5mm}
\section{Deep Learning Based Port Selection}\label{S3}
\vspace{-.4mm}
\subsection{Motivation and Problem Formulation}\label{S3.1}
\vspace{-.4mm}

Since the transmit power at UE side is quite limited, the measured uplink CSI usually suffers from a low SNR, which incurs the inaccuracy of port selection and degrades the performance of the Type-II codebook.
Fortunately, the CSI matrix possesses the sparse structure in the angular-delay domain \cite{ref2}.
Due to the power leakage, the coefficients of different ports are correlated \cite{ref49}, which can be used for enhancing the selection accuracy.
Nevertheless, the sparse structure in the multipath environment is fairly complicated and hard to be accurately extracted by the conventional methods.
Inspired by the strong adaptive fitting capabilities, deep learning is adopted to capture the sparse features for accurately selecting the optimal ports.

Specifically, let $\mathcal{P}$ denote the index set of selected angular-delay-domain ports.
Then, the port selection function $f(\cdot)$ can be written as
\vspace{.8mm}
\begin{equation}\label{eq9}
\mathcal{P} = f\left(\bm{H}_\text{UL} \right).
\vspace{-1.5mm}
\end{equation}
For the selection of single optimal port, the traditional studies usually formulate this problem as a multi-classification task, where each possible selection is viewed as a class \cite{ref50}.
However, the output $\mathcal{P}$ is a set that contains $\binom{N_\text{Tx} M}{P}$ possible combinations, hence it is not feasible to establish a multi-classification for solving (\ref{eq9}).
To address this problem, (\ref{eq9}) is decomposed into $N_\text{Tx} M$ subproblems, where the $n$-th subproblem judges whether the $n$-th port is selected.
Consequently, it is transformed into a multi-label classification task with $N_\text{Tx} M$ labels as below
\vspace{-2mm}
\begin{align}
& I_n = f_n\left(\bm{H}_\text{UL} \right), n \in \left\{ 1, 2, ..., N_\text{Tx} M \right\}, \label{eq10_basic} \\
\vspace{-1mm}
& \text{s.t.}~\sum_{n=1}^{N_\text{Tx} M} I_n = P, \tag{\ref{eq10_basic}{a}} \label{eq10a}
\end{align}
\vspace{-3.5mm}\newline{where the $n$-th label $I_n$ satisfies $I_n = 1$ if the $n$-th port is selected (i.e., within the $P$ ports with largest power) and otherwise $I_n = 0$, while $f_n(\cdot)$ denotes the corresponding selection function.
By tackling the binary-classification tasks for each label under the constraint that the total number of positive classifications equals to $P$, the selected port set $\mathcal{P}$ can be acquired as the indices with positive classifications.}

\vspace{-1mm}
\subsection{Proposed Model Design with Focal Loss}\label{S3.2}
\vspace{-.5mm}

Because the selection of different ports is correlated, a unified deep learning model is constructed to extract the CSI features for all subproblems, which exhibits $N_\text{Tx} M$ binary outputs as the classification result.
In consistent with the existing studies \cite{ref7,ref8,ref9,ref10,ref11}, the convolutional neural network (CNN) is adopted as the backbone architecture for port selection. Specifically, our model design consists of three components, as shown in Fig.~1(a).

\textbf{1) Preprocessing Module}: To better capture the sparse features, the uplink CSI matrix $\bm{H}_\text{UL}$ is firstly transformed to the angular-delay-domain CSI $\widetilde{\bm{H}}_\text{UL}$ as the model input, then the maximum amplitude of the elements in $\widetilde{\bm{H}}_\text{UL}$ is normalized into $1$.
Besides, the normalized CSI $\widetilde{\bm{H}}_\text{UL}^\text{N}$ is divided into two real-valued feature channels corresponding to its real and imaginary parts $\big\{ \Re \big( \widetilde{\bm{H}}_\text{UL}^\text{N} \big), \Im \big( \widetilde{\bm{H}}_\text{UL}^\text{N} \big) \big\}$, which are fed into the following convolutional module.

\textbf{2) Convolutional module}: Multiple convolutional blocks are adopted to learn the sparse features, where each block is made up of a convolutional layer, a batch normalization (BatchNorm) layer and a LeakyReLU activation function in order, as depicted in Fig.~1(b).
After the ultimate convolutional block, a pooling layer is used to downsample each feature channel to a scalar.

In order to enhance the accuracy of port selection, the circular padding is proposed to be exploited in the convolutional layers.
Specifically, the two opposite edges of the angular-delay-domain CSI are actually `adjacent' because of the cyclic property of DFT bases.
However, the broadly applied zero padding ignores this property and directly adds zeros at the edges, which buries their correlations and results in inadequate learning of the sparse structure.
By contrast, the circular padding adopts the opposite edge as the padding contents, so that the correlations between the edges can be fully reserved.

\begin{figure}[tp!]
\begin{center}
\includegraphics[width=.5\textwidth]{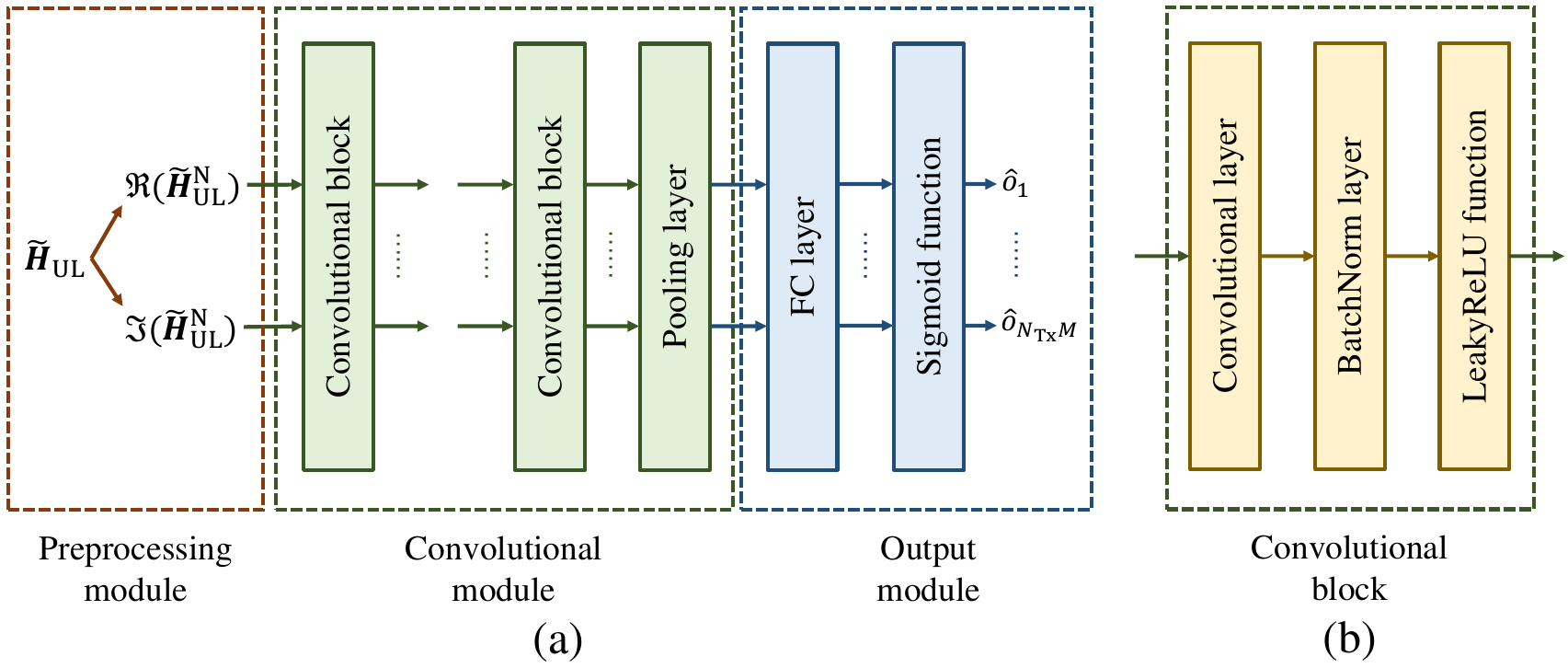}
\end{center}
\vspace{-6mm}
\caption{Illustrations of (a) proposed deep learning model for port selection and (b) convolutional block.}
\label{fig1} 
\vspace{-6mm}
\end{figure}


\textbf{3) Output module}:
A fully-connected (FC) layer is introduced after the pooling layer to implement the transformation from the extracted features to the angular-delay-domain ports, followed by a sigmoid function to normalize the range of each output $\widehat{o}_{n}$ into $(0, 1)$.
Since the output vector $\widehat{\bm{o}} = \left[ \widehat{o}_{1} ~ \widehat{o}_{2} ~ ... ~ \widehat{o}_{N_\text{Tx} M} \right]^\text{T}$ illustrates the priority of port selection, the $P$ ports with highest outputs are chosen as the final result.



To optimize the model parameters, the binary cross entropy (BCE) loss is utilized for each port selection, and the loss of the whole model can be written as
\vspace{-1.4mm}
\begin{equation}\label{eq14}
\text{loss} = - \frac{1}{N_\text{Tx}M}\sum\limits_{n=1}^{N_\text{Tx}M} \big[ I_n \log\widehat{o}_n + (1 - I_n) \log\left(1 - \widehat{o}_n\right) \big].
\vspace{-1mm}
\end{equation}

Nevertheless, directly using (\ref{eq14}) to train the model may not achieve an adequately accurate port selection for two reasons.
Firstly, since the number of selected ports $P$ is generally much smaller than that of all ports $N_\text{Tx} M$, the optimization process would mainly pay attention to the discarded ports and underrate the selected ports with more importance.
Secondly, it is more difficult to accurately select the port with smaller power than that with larger power from all the $P$ optimal ports, whereas (\ref{eq14}) provides the equal weight for both simple and difficult ports.
To address these issues, the focal loss \cite{ref53} is harnessed to balance the training samples for enhancing the selection accuracy.
Specifically, the focal loss firstly adds the weights to balance the losses of positive class (i.e., the selected ports) and negative class (i.e., the discarded ports), where the weight ratio can be empirically set as the reciprocal of the ratio of sample numbers.
Then, the positive and negative labels are separately scaled by the modulating factors $\left\{(1 - \widehat{o}_n)^\gamma, \widehat{o}_n^\gamma\right\}$ with a tunable focusing
parameter $\gamma \geq 0$, which could aid the loss function to concentrate more on the difficult ports.
For example, the positive port with output $\widehat{o}_n$ near $0$ is regarded as a difficult port and would have a relatively large modulating factor $(1 - \widehat{o}_n)^\gamma$.
In summary, the BCE loss of our model improved by the focal loss (FL) is expressed as
\vspace{-2.3mm}
\begin{align}\label{eq15}
 \text{loss}_\text{FL} = & - \frac{1}{N_\text{Tx}M}\sum\limits_{n=1}^{N_\text{Tx}M} \Big[ \frac{N_\text{Tx}M - P}{N_\text{Tx}M} (1 - \widehat{o}_n)^\gamma I_n \log\widehat{o}_n + \nonumber \\
  & \frac{P}{N_\text{Tx}M} \widehat{o}_n^\gamma (1 - I_n) \log\left(1 - \widehat{o}_n\right) \Big].
\end{align}
\vspace{-3.7mm}

The deployment of our proposed model consists of two stages, training and predicting.
At the training stage, training data are collected to optimize the model, where each sample comprises the noisy uplink CSI matrix as the input and the indices of $P$ optimal ports as the label.
The labels can be obtained based on the high-SNR angular-delay-domain CSI matrices estimated by long pilot sequences.
Once the model is well trained with adequate data, it switches to the predicting stage, where the BS adopts the estimated uplink CSI matrix to select the strongest ports based on the model for achieving high accuracy.

\vspace{-1.3mm}
\section{Deep Learning Based CSI Reconstruction}\label{S4}
\vspace{-.6mm}
\subsection{Motivation and Problem Formulation}\label{S4.1}
\vspace{-.6mm}
Assuming perfect port selection, the quantization errors of feeding back the port coefficients of downlink CSI become another performance bottleneck of the R17 Type-II codebook.
To tackle this problem, most of the existing deep learning based works utilize the AE to implement the feedback \cite{ref19,ref20,ref21,ref22}.
Relying on the determined CSI structure, these traditional methods can sufficiently exploit the sparse features reflected by the correlations between different CSI elements to improve the feedback performance.
Nevertheless, this advantage cannot be well leveraged for the R17 Type-II codebook, because the selected angular-delay-domain ports for the UEs at diverse locations are quite different.
Considering that UE is not aware of the corresponding bases of the measured port coefficients, the coefficient vector for feedback does not have a determined sparse structure in the view of UE.
More seriously, deep learning suffers from the low-rank bias \cite{ref24, ref25}, i.e., it tends to extract the low-rank sparse features and is not good at handling irregular non-sparse data, which severely degrades the performance of compressing the coefficient vector.
In addition, the considerable calculational overhead of deep learning based CSI compression at UE side is a vital concern for practical deployments \cite{ref11}.

To solve the problems above, we propose to apply the quantization process in the R17 Type-II codebook to compress the coefficient vector of selected ports for feedback, and focus on the CSI reconstruction at BS side.
Because the BS is aware of the angular-delay-domain basis for each feedback coefficient, it can place all the coefficients on their corresponding ports, which forms the reconstructed CSI by the Type-II codebook according to (\ref{eq8}).
It is clear that the sparse structure in this reconstructed CSI is fully reserved, thus deep learning can be exploited to extract the sparse features for enhancing the reconstruction accuracy.

Specifically, let $\bm{H}_{\text{DL(p)}} \in \mathbb{C}^{N_\text{Tx} \times M}$ represent the perfect reconstructed downlink CSI, which contains the coefficients of all $N_\text{Tx} M$ ports without quantization errors.
Consequently, the deep learning based CSI reconstruction problem can be formulated as a regression task as below
\vspace{-1.7mm}
\begin{equation}\label{eq16}
\bm{H}_{\text{DL(p)}} = g\left( \bm{H}_{\text{DL(TypeII)}} \right),
\vspace{-1.7mm}
\end{equation}
where $g(\cdot)$ denotes the regression function.
To better capture the correlations of port coefficients, we transform the reconstruction problem (\ref{eq16}) to the angular-delay domain, i.e., $ \widetilde{\bm{H}}_{\text{DL(p)}} = g_0 \big( \widetilde{\bm{H}}_{\text{DL(TypeII)}} \big) $ with its corresponding regression function $g_0 (\cdot)$.

\vspace{-1.5mm}
\subsection{Proposed Model Design with Weighted Shortcut}\label{S4.2}
\vspace{-.4mm}

Similar to Subsection \ref{S3.2}, CNN is adopted as the backbone architecture to implement the CSI reconstruction. The key difference lies in the output module, so we only detail this module in the subsection for brevity.


Considering that the selected ports usually occupy the vast majority of CSI power and the quantized port coefficients are close to their original values, the reconstructed angular-delay-domain CSI based on the Type-II codebook $\widetilde{\bm{H}}_{\text{DL(TypeII)}}$ has high similarity to its perfect counterpart $\widetilde{\bm{H}}_{\text{DL(p)}}$.
Therefore, to alleviate the burden of deep learning, we propose to regress the CSI difference between $\widetilde{\bm{H}}_{\text{DL(TypeII)}}$ and $\widetilde{\bm{H}}_{\text{DL(p)}}$, i.e., $ \widetilde{\bm{H}}_{\text{DL}(\Delta)} = \widetilde{\bm{H}}_{\text{DL(p)}} - \widetilde{\bm{H}}_{\text{DL(TypeII)}}$, and a shortcut module is introduced to add the input $\widetilde{\bm{H}}_{\text{DL(TypeII)}}$ for acquiring the final output.
Besides, because the amplitude of $\widetilde{\bm{H}}_{\text{DL}(\Delta)}$ is generally so small that deep learning may be hard to accurately fit, the output of deep learning is multiplied by a weighting coefficient $w < 1$, which can be empirically set around the average amplitude of $\widetilde{\bm{H}}_{\text{DL}(\Delta)}$.
In the corresponding output module, the FC layer transforms the feature vector from the pooling layer into the vectorized CSI difference, followed by the reshaping and weighting operations to obtain the ultimate reconstructed CSI difference.



The mean squared error loss is applied for model optimization, which can be written as
\vspace{-1.5mm}
\begin{equation}\label{eq17}
\text{loss} = \frac{1}{N_\text{Tx} M} \Big|\Big| \widetilde{\bm{H}}_{\text{DL(p)}} - \widehat{\widetilde{\bm{H}}}_{\text{DL(p)}} \Big|\Big|_\text{F}^2,
\vspace{-3mm}
\end{equation}
where $\widehat{\widetilde{\bm{H}}}_{\text{DL(p)}}$ denotes the reconstructed perfect angular-delay-domain CSI by deep learning.
The deployment of the CSI reconstruction model is similar to port selection. To support model training, the almost perfect CSI label can be acquired by channel estimation with long pilot sequences.

\vspace{-2mm}
\section{Simulation Study}\label{S5}
\vspace{-.5mm}
\subsection{System Setup}\label{S5.1}
\vspace{-.5mm}

In the simulations, a single-BS FDD system with $K=5$ UEs is considered, where the 3GPP 38.901 channel model for urban macro-cell scenarios is utilized to generate the CSI matrices \cite{ref26}.
For simplicity, the BS is assumed to adopt zero-forcing precoding for simultaneously serving all UEs \cite{ref0}.
For the R17 Type-II codebook, unless otherwise stated, the selection of $P = 32$ dominant angular-delay-domain ports from $N_\text{Tx} M = 256$ ports is considered for downlink channel measurement and feedback.
In the port coefficient compression at UE side, the wideband amplitude coefficient is quantized by $Q_\text{w} = 5$ bits, while $Q_\text{n,a}=3$-bit and $Q_\text{n,p} = 4$-bit quantizations are adopted for narrowband amplitude and phase coefficients \cite{ref18}.
The detailed parameters are listed in Table~\ref{tab1}.

The specific structures of our deep learning based port selection and CSI reconstruction models are shown in Table~\ref{tab2}, where the both models adopt the same structure of the convolutional module for simplicity.
Concretely, $f_\text{i}$ and $f_\text{o}$ represent the numbers of input and output feature channels.
Besides, the two-dimensional convolutional kernel $(3,3)$ and circular padding $(1,1)$ as well as the LeakyReLU function with negative-axis slope $0.1$ are shared by all convolutional blocks (CBs), thus omitted in Table~\ref{tab2}.
For port selection, the focusing parameter in the focal loss is set to $\gamma = 2$ \cite{ref53}.

\begin{table}[tp!] 
\vspace*{-1mm}
\caption{System parameters.}
\label{tab1}
\vspace*{-5mm}
\begin{center}
\setlength{\tabcolsep}{2mm}
\begin{tabular}{ccccccc}
\toprule[0.8pt]
Parameters                                                                      & Values    \\ \toprule[0.8pt]
Uplink/downlink center frequency    $f_\text{c,UL}/f_\text{c,DL}$               & $3.4/3.5$ GHz \\
Cell radius $r$                                                                 & $250$ m  \\
Downlink transmit power    $P_\text{Tx}$                                        & $35$ dBm  \\
Noise factor      $N_{\text{F}}$                                                & $5$ dB \\
Number of BS antennas $N_\text{Tx}$                                             & $32$ \\
Numbers of BS horizonal/vertical antennas $N_\text{h}/N_\text{v}$               & $4/4$ \\
Subcarrier spacing         $f_\text{s}$                                         & $15$ kHz \\
Number of subbands         $M$                                                  & $8$ \\
\toprule[0.8pt]
\end{tabular}
\end{center}
\vspace{-5mm}
\end{table}

{
\begin{table}[tp!] 
\vspace{0mm}
\renewcommand\arraystretch{1.0}
\makeatletter\def\@captype{table}\makeatother\caption{Proposed deep learning structures.}
\label{tab2}
\vspace*{-5mm}
\begin{center}
\begin{tabular}{cccc}
\toprule[0.8pt]
\rule{0pt}{5pt}
\footnotesize{Modules}                                                     & \footnotesize{Layers} & \footnotesize{Parameters} \\
\toprule[0.8pt]
\multirow{3}{*}{\tabincell{c}{Convolutional \\ module}}
& CB 1                   & $f_\text{i}= 2, f_\text{o}= 256,\text{stride}=(1,1)$\\
& CB 2                   & $f_\text{i}= 256, f_\text{o}= 512, \text{stride}=(3,1)$\\
& CB 3--5                   & $f_\text{i}= 512, f_\text{o}= 512, \text{stride}=(3,3)$\\
\toprule[0.8pt]
\multirow{2}{*}{\tabincell{c}{Output module for \\ port selection}}       & FC              & $f_\text{i}= 512, f_\text{o}= N_\text{Tx}M,\text{dropout}=0.3$  \\
                                                                          & Sigmoid              & / \\
\toprule[0.8pt]
\multirow{2}{*}{\tabincell{c}{Output module for \\ CSI reconstruction}}    & FC              & $f_\text{i}= 512, f_\text{o}= N_\text{Tx}M,\text{dropout}=0.1$  \\
                                                                          & Shortcut              & / \\
\toprule[0.8pt]
\end{tabular}
\end{center}
\vspace{-9.5mm}
\end{table}
}

The training dataset containing $102,400$ samples and the validation dataset containing $2,560$ samples are constructed, respectively. The deep learning models are trained with learning rate $3 \times 10^{-6}$ for $200$ epochs in the training stage, where Adam optimizer is used to optimize the model.

\vspace{-2mm}
\subsection{Simulation Results on Port Selection}\label{S5.2}
\vspace{-.7mm}

The performance of port selection is evaluated by two metrics.
The first metric is the normalized CSI power of the selected ports $P_\text{N}$, which can be calculated as
\vspace{-2mm}
\begin{equation}\label{eq19}
P_\text{N} = \frac{ \sum_{p=1}^{P} \big|{\widetilde{\bm{H}}_\text{DL}}[{p}^{\text{(A)}},{p}^{\text{(D)}}]\big|^2 }{ \big|\big|\widetilde{\bm{H}}_\text{DL}\big|\big|_\text{F}^2 },
\vspace{-2mm}
\end{equation}
where ${p}^{\text{(A)}}$ and ${p}^{\text{(D)}}$ denote the angular-domain and delay-domain indices of the $p$-th selected port, respectively.
Clearly, higher $P_\text{N}$ indicates more accurate port selection.
The second metric is the average sum rate $R_\text{avg}$ after precoding \cite{ref0}.
For fair comparisons, we apply the standard procedure of the R17 Type-II codebook except for port selection in this subsection.

\begin{figure}[tp!]
\vspace{-1mm}
\begin{center}
\includegraphics[width=.35\textwidth]{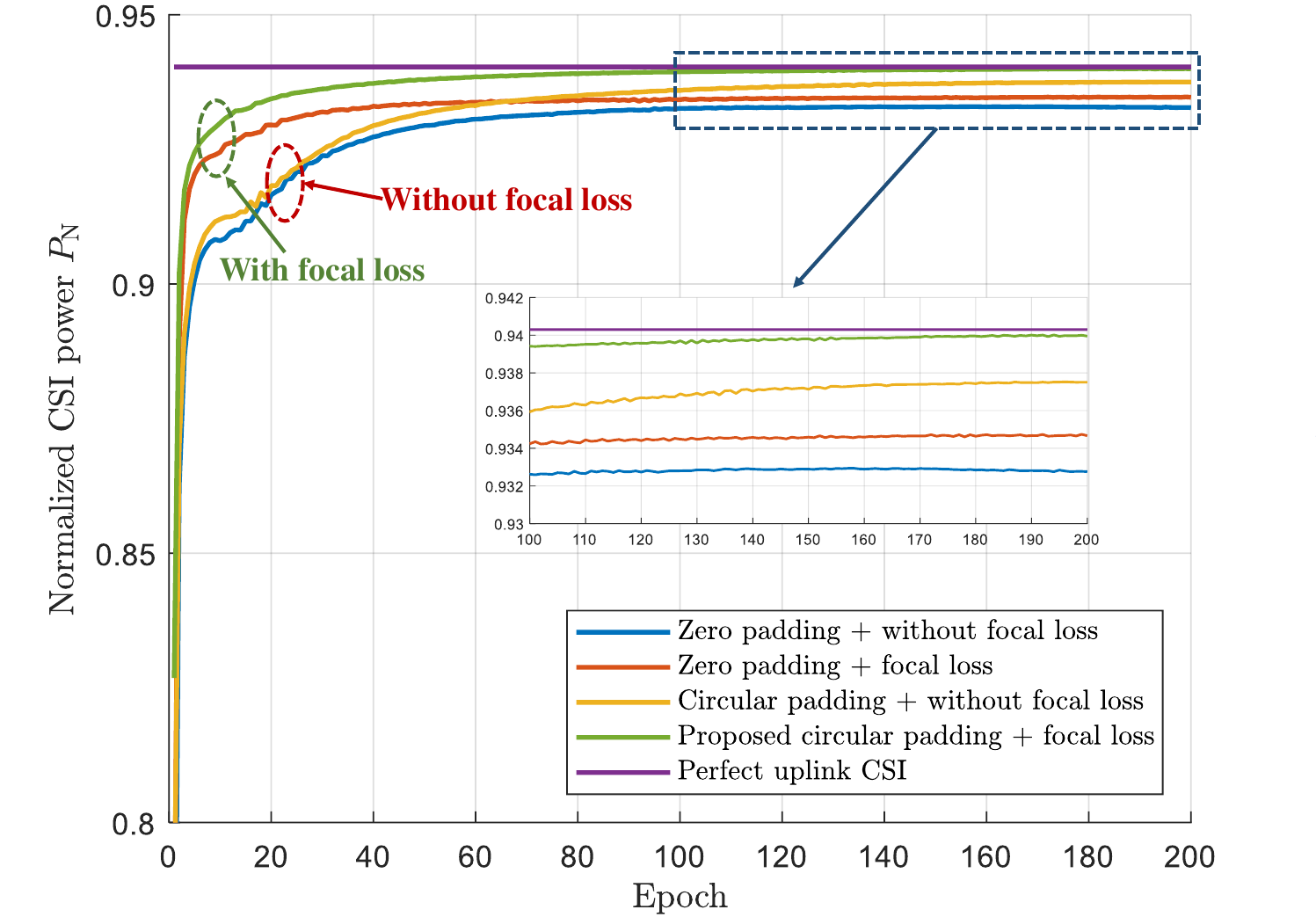}
\end{center}
\vspace{-6mm}
\caption{Convergence performance comparison of different deep learning designs in terms of normalized CSI power, where uplink CSI SNR is $5$~dB.}
\label{fig3}
\vspace{-3mm}
\end{figure}

Firstly, Fig.~\ref{fig3} depicts the convergence performance of our proposed deep learning model in terms of the normalized CSI power $P_\text{N}$ under the uplink CSI $\text{SNR}=5$~dB, where the impact of our adopted circular padding and focal loss is also investigated.
Besides, the $P_\text{N}$ performance of the port selection from the perfect uplink CSI without noise is illustrated as the upper bound.
By comparing different deep learning designs, it can be concluded that both the circular padding and the focal loss are beneficial for achieving higher $P_\text{N}$.
Besides, the models with the focal loss could obtain a significantly faster convergence speed before the $60$-th epoch, since the focal loss can concentrate on difficult samples for accelerating model optimization.
Furthermore, our proposed model design with the circular padding and focal loss surprisingly attains almost the same $P_\text{N}$ performance as the perfect uplink CSI.

\begin{figure}[tp!]
\vspace{0mm}
\begin{center}
\includegraphics[width=.35\textwidth]{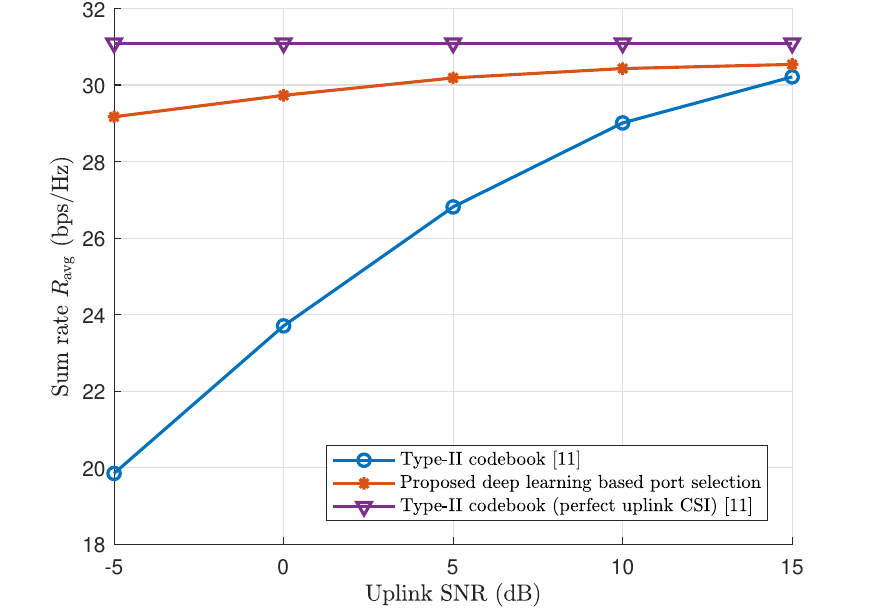}
\end{center}
\vspace{-6mm}
\caption{Sum rate performance comparison for different port selection methods as function of uplink SNR.}
\label{fig4}
\vspace{-7mm}
\end{figure}

Next, the sum rate performance $R_\text{avg}$ of our proposed port selection method and the Type-II codebook \cite{ref18} under different SNRs of uplink CSI are compared in Fig.~\ref{fig4}, where the achievable $R_\text{avg}$ of the uplink CSI without noise is also depicted.
It can be seen that the $R_\text{avg}$ performance of our proposed method surpasses the Type-II codebook in all evaluated SNRs from $-5~\text{dB}$ to $15~\text{dB}$, especially for the severely low-SNR scenarios.
Besides, our proposed method enjoys the more robust $R_\text{avg}$ performance to uplink SNR than the Type-II codebook, and only suffers from $6.2\%$ rate loss compared to the perfect uplink CSI under $-5~\text{dB}$ uplink SNR.

\vspace{-2mm}
\subsection{Simulation Results on CSI Reconstruction}\label{S5.3}
\vspace{-.7mm}

\begin{figure}[tp!]
\vspace{-1mm}
\begin{center}
\includegraphics[width=.34\textwidth]{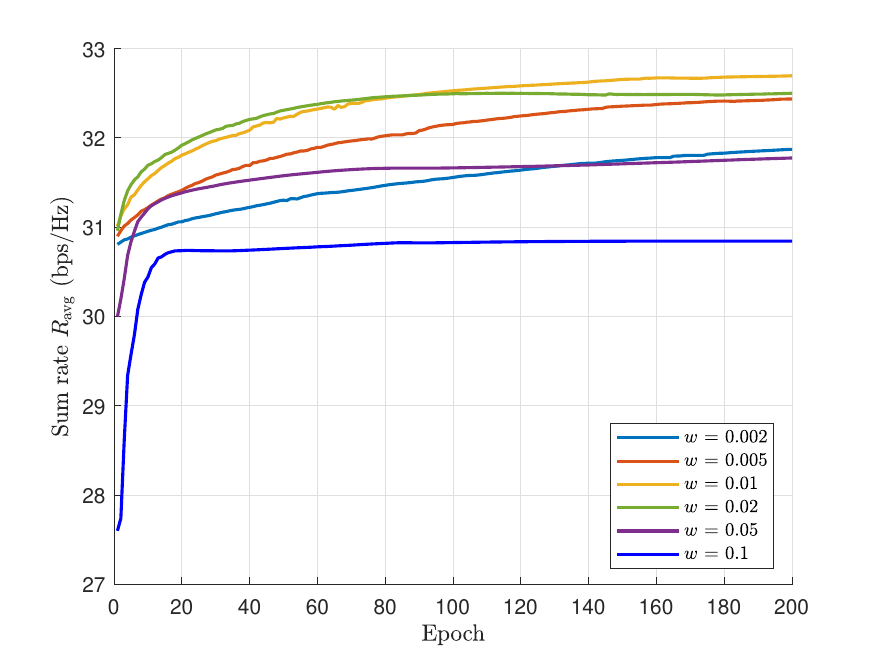}
\end{center}
\vspace{-6mm}
\caption{Sum rate performance comparison under different weighting coefficients, where uplink SNR is $5~\text{dB}$.}
\label{fig5}
\vspace{-4mm}
\end{figure}

Based on the proposed port selection method, we further investigate the CSI reconstruction performance in terms of the sum rate $R_\text{avg}$.
Firstly, the impact of the weighting coefficient in the shortcut module $w$ on the convergence of $R_\text{avg}$ is shown in Fig.~\ref{fig5}, where the SNR of uplink CSI is set to $5$~dB.
Since the average amplitude of the angular-delay-domain CSI difference $\widetilde{\bm{H}}_{\text{DL}(\Delta)}$ is $0.017$ in the training dataset, the empirical value of $w$ is chosen from a nearby range.
It can be seen that the moderate $w \in \{ 0.005,0.01,0.02 \}$ around $0.017$ enjoys high sum rate performance, while both overwhelmingly small and large $w$ could degrade $R_\text{avg}$ owing to the amplitude constraint of batch normalization.
Based on Fig.~\ref{fig5}, we adopt the optimal parameter $w=0.01$ in the following simulations.

%

\begin{figure}[tp!]
\vspace{-0mm}
\begin{center}
\includegraphics[width=.34\textwidth]{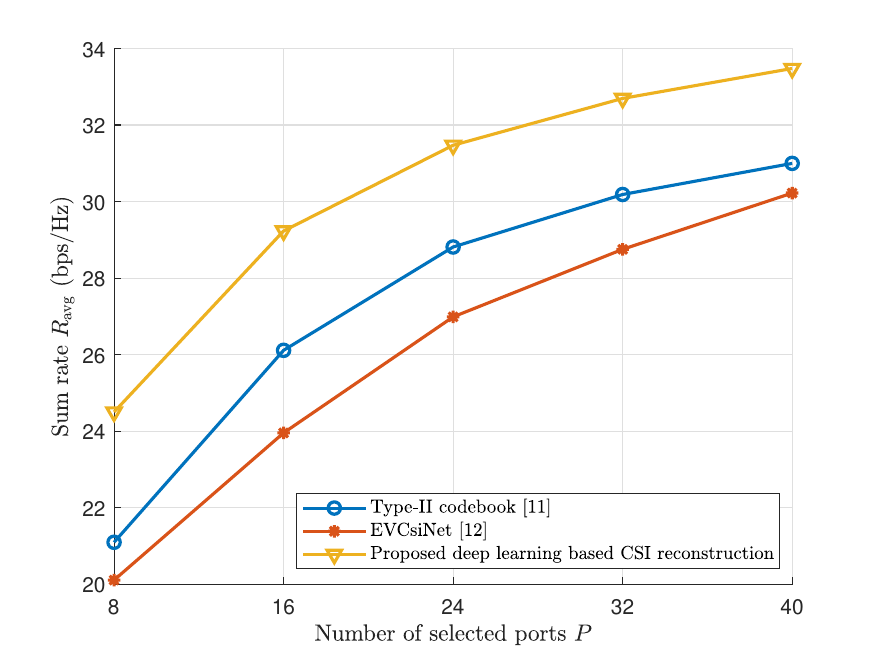}
\end{center}
\vspace{-6mm}
\caption{Sum rate performance comparison for different CSI feedback methods as function of selected port number.}
\label{fig7}
\vspace{-7mm}
\end{figure}

Finally, Fig.~\ref{fig7} compares the sum rate performance $R_\text{avg}$ of different CSI feedback methods as the function of the selected port number $P$, where the standard feedback procedure of the R17 Type-II codebook \cite{ref18} and the AE based EVCsiNet \cite{ref19} are adopted as our baselines.
For a fair comparison, the number of feedback bits in EVCsiNet is equal to the R17 Type-II codebook.
We can see that the $R_\text{avg}$ performance of the conventional EVCsiNet is poorer than the standard Type-II codebook, which validates that deep learning is not expert in compressing the non-sparse vector of port coefficients.
By contrast, our proposed method could achieve higher $R_\text{avg}$ than both baselines, which demonstrates that the proposed method could effectively leverage the sparse structure for improving the reconstruction accuracy.

\vspace{-3mm}
\section{Conclusions}\label{S6}
\vspace{-2mm}
In this paper, we have proposed two new perspectives for improving the CSI feedback performance of the R17 Type-II codebook.
Firstly, we leverage deep learning to accurately select the dominant angular-delay-domain ports from the uplink CSI with low SNRs, where the circular padding and focal loss are harnessed to enhance the selection accuracy.
Secondly, we elaborate that the key limitation of adopting the conventional AE based CSI feedback methods in the R17 Type-II codebook is the deficiency of the sparse structure in the feedback coefficient vector.
Consequently, we adopt deep learning to reconstruct the CSI at BS side for fully exploiting the sparse features, where a weighted shortcut module is utilized to enhance the reconstruction accuracy.
The simulation study has demonstrated that our proposed methods could achieve higher sum rate performance compared to its traditional R17 Type-II codebook and deep learning benchmarks.

\vspace{-1.5mm}
\section*{Acknowledgment}
\vspace{-1mm}

{\small This work was supported in part by the National Natural Science Foundation of China under Grant U22B2057, in part by Guangdong Optical Wireless Communication Engineering and Technology Center, in part by Shenzhen VLC System Key Laboratory, in part by Shenzhen Solving Challenging Technical Problems (No.~JSGG20220831100601002), and in part by Huawei Research Fund. (\emph{Corresponding author: Zhaocheng Wang})}

\vspace{-1mm}
{
}

\end{document}